\documentclass[12pt]{article}
\begin{document}

\begin{titlepage}
\begin{flushright} KUL--TF--95/33 \\
   hep-th/9510195  \\
                   October 1995\\
\end{flushright}

\vfill
\begin{center}
{\LARGE
Background charges and consistent continuous deformations
of $2d$ gravity theories}\\
\vfill
{\Large {F}riedemann {Brandt} $^1$,
{W}alter {Troost} $^{2,3}$\\
\vskip 1.5mm and {A}ntoine {Van Proeyen}
$^{2,4}$} \\
\vfill
Instituut voor Theoretische Fysica
        \\Katholieke Universiteit Leuven
        \\Celestijnenlaan 200D
        \\B--3001 Leuven, Belgium\\
\end{center}
\vfill
\begin{center}
{\bf Abstract}
\end{center}
\begin{quote}
\small
We discuss all background charges and continuous consistent
deformations of standard $2d$ gravity theories with scalar matter fields.
The background charges and those deformations
which change nontrivially the gauge symmetries
are closely linked, and exist only if the target
space has a covariantly constant Killing
vector (a null vector for the
deformations). The deformed actions provide interesting novel $2d$
gravity models. Some of them lead to non-critical string theories.
\vspace{2mm} \vfill \hrule width 3.cm
{\footnotesize
\noindent $^1$ Junior fellow of the research council (DOC) of the
K.U. Leuven;\\
\phantom{$^1$} E--mail : Friedemann.Brandt@fys.kuleuven.ac.be \\
\noindent $^2$ Onderzoeksleider, NFWO, Belgium\\
\noindent $^3$
E--mail : Walter.Troost@fys.kuleuven.ac.be\\
\noindent $^4$
E--mail : Antoine.VanProeyen@fys.kuleuven.ac.be}
\normalsize
\end{quote}
\end{titlepage}

Weyl invariant $2d$ gravity models
have attracted a lot of attention because they
underly various systems of physical interest
(string theory, two-dimensional $\sigma$- and
WZWN-models, conformal field theories).
As these models are gauge theories, many of their properties
can be analysed by BRST techniques. For a class
of such models we computed in \cite{1} (cf. also \cite{1a})
the cohomology groups $H^g({\cal S})$ of the BRST--Slavnov operator ${\cal S}$
in the space of local functionals for all ghost numbers $g$.
This letter is devoted to the discussion of the results
for $H^0({\cal S})$: we will use them to construct all
so-called background charges for these models in the
form described in \cite{leu}
and all continuous deformations of the classical actions
which are consistent in the sense that the deformed actions
are invariant under possibly deformed gauge
transformations \cite{bh}.

It turns out that
both the background charges and those deformations
which change nontrivially the form of the
gauge transformations exist only
for target spaces with a covariantly constant (Killing) vector which
must be a null vector in the case of the deformations. Such models
deserve special attention also for other reasons:
after gauge fixing (e.g. in the conformal gauge) they
have a residual (chiral) Ka\v{c}--Moody symmetry and can thus
provide interesting conformal field theories; they allow to
gauge target space isometries \cite{hs};
and some so-called `chiral null models' provide solutions of the string
equations of motion which are exact to all orders in $\alpha'$ \cite{ht}.

Let us first recall the results of \cite{1} on $H^0({\cal S})$.
The models in question are fixed solely by
the field content and by the gauge symmetries of the classical
theory: the fields of the classical theory are
the $2d$ metric $g_{\alpha\beta}=g_{\beta\alpha}$
($\alpha,\beta\in\{+,-\}$)
and an arbitrary number of Weyl invariant scalar
`matter fields' $X^\mu$
($\mu=1,\ldots,D$); the gauge symmetries are the standard $2d$
diffeomorphisms and Weyl transformations.
The corresponding nilpotent BRST transformations read
\begin{eqnarray}
sg_{\alpha\beta}&=&\xi^\gamma \partial_\gamma g_{\alpha\beta}+
g_{\gamma\beta}\partial_{\alpha} \xi^\gamma
+g_{\alpha\gamma}\partial_{\beta} \xi^\gamma
+c\, g_{\alpha\beta}\ ;\nonumber\\
 s\Phi &=&\xi^\alpha \partial_\alpha \Phi\quad \mbox{for}
\quad \Phi\in\{X^\mu,\xi^\beta,c \}\label{2a}\end{eqnarray}
where the $\xi^\alpha$ are the diffeomorphism ghosts
and $c$ is the Weyl ghost.
As we have shown in \cite{1}, any {\it local} classical action
constructible out of
$g_{\alpha\beta}$ and $X^\mu$ and invariant under these
gauge transformations is, up to boundary terms, of the simple
well-known form
\begin{equation}S_{cl}=
\int\mbox{\small{$\frac 12$}}
\left[ \sqrt{g}\, g^{\alpha\beta}G_{\mu\nu}(X)
+\varepsilon^{\alpha\beta}B_{\mu\nu}(X)\right]
\partial_\alpha X^\mu \cdot \partial_\beta X^\nu
\label{1}\end{equation}
where $g=-\det (g_{\alpha\beta})$ (we assume $g>0$),
$\varepsilon^{+-}=-\varepsilon^{-+}=1$ and
$\int\equiv\int d^2 x$.
Hence, different models are distinguished only through
the functions
$G_{\mu\nu}(X)$ and $B_{\mu\nu}(X)$ occurring in (\ref{1})
which are symmetric resp. antisymmetric in $\mu,\nu$
($G_{\mu\nu}$ is not assumed to be necessarily invertible).
As the  gauge algebra is
closed and irreducible, the
(minimal) solution $S$ of the master equation \cite{bv}
corresponding to (\ref{1}) reads
\begin{equation} S=S_{cl}-\int (s\,\Phi^A)\, \Phi^*_A\ ; \quad
\{\Phi^A\}=\{g_{\alpha\beta},X^\mu,\xi^\alpha,c\}
\label{4}\end{equation}
where $\Phi^*_A$ is the antifield of $\Phi^A$.
The BRST--Slavnov operator ${\cal S}$ whose cohomology was computed
in \cite{1} is defined on any
function(al) $F$ of the fields and antifields
through the antibracket \cite{bv} of $S$ and $F$,
${\cal S} F:= (S,F)$.
${\cal S}$ is a nilpotent antiderivation because $S$ solves
the master equation, $(S,S)=0$.

The results of \cite{1} state now that, for any given
$G_{\mu\nu}$ and $B_{\mu\nu}$, there are at most two
types of nontrivial cohomology classes in $H^0({\cal S})$. Those of
the first type are obvious ones
represented by functionals $W^0_{(0)}$
of the same form as the action (\ref{1}) itself,
\begin{equation}W^0_{(0)}=\int  \mbox{\small{$\frac 12$}}
[\sqrt{g} g^{\alpha\beta}N_{\mu\nu}(X)
+\varepsilon^{\alpha\beta}F_{\mu\nu}(X)]
\partial_\alpha X^\mu \cdot \partial_\beta X^\nu.
\label{1a}\end{equation}
$W^0_{(0)}$ is an ${\cal S}$-cocycle for any choice of
$N_{\mu\nu}$ and $F_{\mu\nu}$ and cohomologically nontrivial unless
\begin{equation} N_{\mu\nu}={\cal L}_H G_{\mu\nu} \ \wedge
\ F_{\mu\nu}={\cal L}_H B_{\mu\nu}+2\partial_{[\mu}B_{\nu]}
\label{trivial}\end{equation}
for some $H^\mu(X)$ and $B_\mu(X)$
(${\cal L}_H$ denotes the  Lie
derivative along $H^\mu$ in the target space).

The representatives of $H^0({\cal S})$ of the second type
depend nontrivially on the antifields. They
exist only if $G_{\mu\nu}$ and $B_{\mu\nu}$ allow
`covariantly constant target space vectors' $\zeta_{+}^\mu(X)$
or $\zeta_{-}^\mu(X)$, solving
\begin{equation}
\partial_\mu \zeta_{\nu\pm} -(\Gamma_{\mu\nu,\rho}
\pm \mbox{\small{$\frac 12$}}H_{\mu\nu\rho})\, \zeta^\rho_\pm=0
\label{16}\end{equation}
where $\zeta_{\mu\pm}=G_{\mu\rho}\zeta^\rho_{\pm}$ and
\begin{eqnarray}
\Gamma_{\mu\nu,\rho}&=&\mbox{\small{$\frac 12$}}
(\partial_\mu G_{\nu\rho}
+\partial_\nu G_{\mu\rho}-\partial_\rho G_{\mu\nu})\ ;\nonumber\\
H_{\mu\nu\rho}&=&
\partial_\mu B_{\nu\rho}
+\partial_\nu B_{\rho\mu}+\partial_\rho B_{\mu\nu}\ .
\label{17}\end{eqnarray}
Note that the $\zeta_{+}$ and $\zeta_{-}$ are covariantly constant
w.r.t. to connections with
torsion $\pm (1/2)H_{\mu\nu\rho}$ respectively, and that they
are special Killing vectors of the target
space characterized through
$G_{\mu\nu}$ and $B_{\mu\nu}$. Introducing bases
$\{\zeta^\mu_{a^\pm}\}$
for them ($a^\pm=1,\ldots,N^\pm$),
the general solutions of (\ref{16}) can be written as
linear combinations of the form
$\zeta^\mu_{\pm}(X)=v^{a^\pm}\zeta^\mu_{a^\pm}(X)$,
where the $v$'s are constant coefficients.
A corresponding basis for the representatives of $H^0({\cal S})$
of the second type
is provided by the following nontrivial ${\cal S}$-cocycles:
\begin{equation} W^0_{a^\pm}=
\int \left(
X^*_\mu \zeta^\mu_{a^\pm}C^\pm_\pm
-j_{a^\pm}^\mp\partial_\pm h_{\mp\mp}\right)
\label{20}\end{equation}
where
\begin{eqnarray} h_{\pm\pm}&=&g_{\pm\pm}(g_{+-}+\sqrt{g})^{-1}\ ;
\label{21}\\
C^\pm_\pm&=&
(\partial_\pm \xi^\pm+h_{\mp\mp}\partial_\pm\xi^\mp)\ ;\label{22}\\
j_{a^\pm}^\alpha&=&(\sqrt g g^{\alpha\beta}\mp
\varepsilon^{\alpha\beta})\, \zeta_{\mu a^\pm}\partial_\beta X^\mu
\label{23}\end{eqnarray}
with $\zeta_{\mu a^\pm}=G_{\mu\nu}\zeta^\nu_{ a^\pm}$.
We note that $h_{\pm\pm}$ are `Beltrami variables', that
the ghost combinations $C^\pm_\pm$ satisfy $sC^\pm_\pm=\xi^\alpha
\partial_\alpha C^\pm_\pm$, and
that the $j_{a^\pm}^\alpha$ are the Noether currents
corresponding to the $\zeta^\mu_{a^\pm}$
(each $\zeta^\mu_{a^\pm}$ generates a rigid symmetry of
$S_{cl}$, cf. \cite{1}). These currents become chiral
in the conformal gauge ($g_{+-}=\sqrt{g}$, $g_{\pm\pm}=0$).
Eq. (\ref{20}) is reminiscent of similar BRST cocycles
in Yang--Mills and Einstein--Yang--Mills theories with $U(1)$-factors
found in \cite{bbh2,bbh3}. The role of the rigid symmetries
in the BRST cohomology was uncovered
in \cite{bbh1}.

We consider now a general linear combination of the ${\cal S}$-cocycles
(\ref{20}) with arbitrary constant coefficients,
\begin{equation} M^0_v=  v^{a^+}W^0_{a^+}+v^{a^-}W^0_{a^-}\ .
\label{24}\end{equation}
Since $M^0_v$ is an ${\cal S}$-cocycle with ghost number 0, the
antibracket $(M^0_v,M^0_v)$
is automatically an ${\cal S}$-cocycle with ghost number one, thanks
to the Jacobi identity for the antibracket.
Hence, $(M^0_v,M^0_v)$ must
be  a linear combination of nontrivial representatives
of $H^1({\cal S})$ modulo an ${\cal S}$-coboundary. Using
properties of the $\zeta$'s derived in
appendix D of \cite{1}, one finds indeed
\begin{eqnarray}  \left(M^0_v,M^0_v\right)=
2P_{++}(v)\, W^1_+-2P_{--}(v)\, W^1_-\nonumber\\
-2\left(S,W^0_{v++}+W^0_{v--}+W^0_{v+-}\right)
\label{25}\end{eqnarray}
where
$W^1_+$ and $W^1_-$ are nontrivial and cohomologically
inequivalent representatives of $H^1({\cal S})$ representing
the potential matter field
independent anomalies \cite{1},
\begin{equation} W_\pm^1=\mp 2
\int (\xi^\pm+h_{\mp\mp}\xi^\mp)(\partial_\pm)^3 h_{\mp\mp}\ ,
\label{27}\end{equation}
and $P_{\pm\pm}(v)$ are constant quadratic forms in the
$v$'s:
\begin{equation}
P_{\pm\pm}(v)=\zeta^\mu_\pm \zeta_{\mu\pm}=constant\ ;\quad
\zeta^\mu_\pm=v^{a^\pm}\zeta^{\mu}_{a^\pm}\ .
\label{26}\end{equation}
(The constancy of these inner products follows from
(\ref{16})). The $W^0$'s occurring in (\ref{25}) are the functionals
\begin{eqnarray} W^0_{v\pm\pm}&=&-P_{\pm\pm}(v)\int
\frac {h_{\pm\pm}(\partial_\pm
 h_{\mp\mp})^2}{1-h_{++}h_{--}}\ ;  \label{28}\\
W^0_{v+-}&=&\int
\frac{2P_{+-}(v,X)\,\partial_+ h_{--}\cdot
\partial_- h_{++}}{1-h_{++}h_{--}}
\label{29}\end{eqnarray}
where $P_{+-}(v,X)$ is the inner product of $\zeta_+^\mu$ and
$\zeta_-^\mu$ which is in general not constant, contrary to (\ref{26}), but
still annihilated by the operators $\zeta^\mu_\pm\partial_\mu$ (see
Eq. (D.21) of \cite{1}),
\begin{equation} P_{+-}(v,X)=\zeta^\mu_+ \zeta_{\mu -}\ ;\quad
\zeta^\mu_\pm\partial_\mu P_{+-}(v,X)=0.\label{zetadP}
\end{equation}

We will now discuss (\ref{25}) and show
that it leads to background charges and continuous
deformations.
It is straightforward to verify that
\begin{equation} S_\tau (v)=S+\tau M^0_{v}+
\tau^2\left(W^0_{v++}+W^0_{v--}+W^0_{v+-}\right) \label{33}
\end{equation}
satisfies for any value of $\tau$
\begin{equation} \left(S_\tau(v),S_\tau(v)\right)=2\tau^2
\left( P_{++}(v) W_+^1- P_{--}(v)W_-^1 \right) \ ,
\label{34}\end{equation}
which holds at $\tau=0$ by construction of $S$, linear in $\tau$
by definition of $M^0_v$, quadratic in $\tau$
due to (\ref{25}), at $\tau^3$ by (\ref{zetadP}) and at
$\tau^4$ because the functionals (\ref{28}) and (\ref{29}) do
not depend on antifields. We use this
result in two different ways:

a) Suppose that
$\hbar\left( a_+W_+^1+a_-W_-^1\right) $
occurs as a one-loop anomaly of the theory.
Suppose further that there are coefficients $\bar v$ satisfying
$2P_{\pm\pm}(\bar v)=\mp a_\pm$. According to (\ref{25}) we can
then formally cancel this anomaly (up to
terms of higher order in $\hbar$) by using as action
$S_{\sqrt{\hbar}}(\bar v)$. Thus $M^0_{\bar v}$ represents  a
background charge, cf. \cite{leu}.
In particular one may cancel in this way the
famous Weyl anomaly whose coefficient is proportional to $(D-26)$.
Namely up to the coefficient this anomaly is represented by
$\int c\sqrt g R$ which is
cohomologically equivalent to $ W^1_+-W^1_-$.

b) The second application
aims to construct a continuous
deformation of the classical action. This is applicable for target
spaces with a non--positive definite metric $G_{\mu\nu}(X)$.
Suppose that a set of $\hat v$'s solves nontrivially
\begin{equation} P_{++}(\hat v)=P_{--}(\hat v)=0.\label{31}\end{equation}
This requires the existence of a
covariantly constant null vector
$\hat\zeta^\mu_+\equiv \hat v^{a^+}\zeta^\mu_{a^+}$ or
$\hat\zeta^\mu_-\equiv \hat v^{a^-}\zeta^\mu_{a^-}$.
Then (\ref{34}) says  that $S_\tau(\hat v)$
satisfies the master equation for any value of $\tau$ and is thus
a consistent continuous deformation of
$S$ with deformation parameter $\tau$.
In fact it is a nontrivial one, i.e. it
cannot be obtained from (\ref{4}) through mere local field
redefinitions because $M^0_{\hat v}$ is a nontrivial
${\cal S}$-cocycle.
Henceforth $S_\tau$ will always denote
$S_\tau(\hat v)$. Let us now briefly discuss it.
It is of the form
\begin{equation} S_\tau=S_{cl,\tau}
-\int  (s_\tau\Phi^A)\, \Phi^*_A
\label{35}\end{equation}
where $S_{cl,\tau}$ is the antifield independent part of $ S_\tau$.
This part is the deformation of the classical action (\ref{1}).
Since (\ref{31}) implies $W^0_{\hat v\pm\pm}=0$,
$S_{cl,\tau}$ is explicitly given by
\begin{eqnarray}
S_{cl,\tau}&=&S_{cl}
+ 2\int  \frac{\tau\, ({\cal L}^++{\cal L}^-)+\tau^2
{\cal L}^{+-}}{1-h_{++}h_{--}}\ ;\nonumber\\
{\cal L}^\pm&=&-\partial_\pm h_{\mp\mp} \cdot\hat\zeta_{\mu \pm}(X)
(\partial_\pm-h_{\pm\pm}\partial_\mp) X^\mu \ ;\nonumber\\
{\cal L}^{+-}& =&\hat\zeta^\mu_+(X)\hat\zeta_{\mu -}(X)\,
\partial_+ h_{--}\cdot \partial_- h_{++}
\label{36}\end{eqnarray}
where we rewrote (\ref{23}) in terms of the Beltrami variables.
Note that $S_{cl,\tau}$ is in general {\it not}
`left-right symmetric', contrary to $S_{cl}$
(left-right symmetry is obtained only for the choice
$\hat\zeta^\mu_+=\hat\zeta^\mu_-$).
(\ref{34}) implies that $S_{cl,\tau}$
is invariant under deformed BRST transformations
(and corresponding deformed gauge transformations) given by
\begin{eqnarray} s_\tau X^\mu&=&\xi^\alpha \partial_\alpha X^\mu+
\tau\, (C^+_+\hat\zeta^\mu_++C^-_-\hat\zeta^\mu_-)\ ;\label{37}\\
s_\tau\Phi&=&s\Phi\quad \mbox{for}\quad
\Phi\in\{g_{\alpha\beta},\xi^\alpha,c\}\ .\label{38}\end{eqnarray}
The fact that (\ref{35}) is linear in the antifields
indicates that the deformed gauge transformations still have
a closed algebra. In fact this algebra agrees
with the original one, as the BRST transformations
of the ghosts remain undeformed, cf. (\ref{38}).
Hence,  even though the gauge transformations themselves
get nontrivially deformed, their commutation relations do not change
(one may check this using
$\hat\zeta^\nu_+\partial_\nu\hat\zeta^\mu_-
-\hat\zeta^\nu_-\partial_\nu\hat \zeta^\mu_+=0$, cf. appendix D of
\cite{1}).

In order to make contact with results known in the literature
and to give some insight into the physical content of the
deformed models, we will now specialise to the case
\begin{equation} \det (G_{\mu\nu})\neq 0,\quad B_{\mu\nu}=0.
\label{sc}\end{equation}
In this case we can obviously choose
$\zeta^\mu_{a^+} =\zeta^\mu_{a^-}$.
Nevertheless the deformed actions (\ref{36}) will
not be left-right symmetric unless we choose also $\hat v^{a^+}=
\hat v^{a^-}$ which yields
\begin{equation} \hat\zeta^\mu_+=
\hat\zeta^\mu_-\equiv\hat\zeta^\mu .\label{40}\end{equation}
Let us now analyse the deformed theory in some more detail for
this special choice. Recall that we must impose
$\hat\zeta^\mu\hat\zeta_\mu=0$, cf. (\ref{31}).
This implies due to (\ref{40}) that the part ${\cal L}^{+-}$
of $S_{cl,\tau}$ vanishes. (\ref{16})
implies in this case
\begin{eqnarray}
& &\partial_{[\mu}\hat\zeta_{\nu]}=0\ \Leftrightarrow
\ \hat\zeta_\mu(X)=\partial_\mu \Lambda(X)
\quad (D\neq 1)\label{41}\\
&\wedge&\quad \partial_\mu\partial_\nu\Lambda-
{\Gamma_{\mu\nu}}^\rho\partial_\rho\Lambda=0.
\label{42}\end{eqnarray}
Using partial integrations (and Eq. (A.12) of \cite{1}) one
straightforwardly verifies that (\ref{36}) reduces in the
special case (\ref{sc}) and for the choice (\ref{40}) to
\begin{eqnarray} S_{cl,\tau}
= \int\sqrt{g}
\left[ \mbox{\small{$\frac 12$}} g^{\alpha\beta}G_{\mu\nu}(X)
\partial_\alpha X^\mu \cdot \partial_\beta X^\nu\right.\nonumber\\
\left.-\tau\Lambda(X) R-\tau g^{\alpha\beta}\partial_\alpha L\cdot
\partial_\beta \Lambda(X)\right]\label{44} \end{eqnarray}
where $R$ is the $2d$ Riemann curvature scalar constructed out of
$g_{\alpha\beta}$, and $L$ is the useful quantity
\begin{equation} L=\ln \frac{\sqrt g}{1-h_{++}h_{--}}
=\ln \frac{\sqrt g+g_{+-}}{2}\ .\label{47a}\end{equation}
One may now verify
that the local field redefinition
\begin{equation} Y^\mu(X,L,\tau)=\exp [-\tau L\,
\hat\zeta^{\nu}(X)\partial/\partial X^\nu]\,  X^\mu
\label{46}\end{equation}
casts (\ref{44}) in a simpler form which is
regular even for $g_{+-}+\sqrt g=0$,
\begin{equation}
S_{cl,\tau}
= \int\sqrt{g}
\left[ \mbox{\small{$\frac 12$}} g^{\alpha\beta}G_{\mu\nu}(Y)
\partial_\alpha Y^\mu \cdot \partial_\beta Y^\nu
-\tau\Lambda(Y) R\right]
\label{regact}\end{equation}
where $G_{\mu\nu}$ and $\Lambda$ are the {\it same} functions
as in (\ref{44}) (but note that their arguments are now the
$Y$'s instead of the $X$'s). To obtain this result, one uses
\begin{eqnarray}
&&\exp [-\tau L\,
\hat\zeta^{\nu}(X)\partial/\partial X^\nu]\,  f(X)=f(Y)\nonumber\\
& &\Rightarrow\quad\hat\zeta^{\nu}(X)(\partial Y^\mu/\partial X^\nu)
=\hat\zeta^{\mu}(Y),\label{42b}
\end{eqnarray}
for any function $f(X)$,
and further that
\begin{equation}
G_{\mu\nu}(X)\frac{\partial X^\mu}{\partial Y^\rho}
\frac{\partial X^\nu}{\partial Y^\sigma}=G_{\rho\sigma}(Y)\ ;
\quad\Lambda(X)=\Lambda(Y) \ ,
\end{equation}
where the first identity holds because $\hat \zeta$ is a Killing vector
and the second one because it is a null vector.
The deformed BRST transformation of $Y^\mu$ reads
\begin{equation} s_\tau Y^\mu=\xi^\alpha\partial _\alpha Y^\mu-\tau\, c\,
\hat\zeta^\mu(Y) \label{48}\end{equation}
which follows also immediately from (\ref{46}) due to (\ref{42b}) and
\begin{equation}
s_\tau L=sL=\xi^\alpha\partial_\alpha L+c+C^+_++C^-_-\ .
\end{equation}

Finally we show that in a model described at classical level by a
deformed action (\ref{regact}) one may
cancel the matter field independent Weyl anomaly by a
counterterm. To that end we
choose a function $\varphi(X)$ such that
\begin{equation}
 \hat\zeta^\mu(X)\, \partial_\mu\varphi(X)=1.
\label{50}\end{equation}
Such a function $\varphi$ always exists. This is easily
seen in a target space
parametrization chosen such that $\hat\zeta^\mu$ has the components
$(1,0,\ldots,0)$ (i.e. the integral curves of $\hat\zeta^\mu$ are
chosen as first coordinate lines in that parametrization).
Then (\ref{50}) requires
$\varphi(X)=X^1+f(X^2,\ldots,X^D)$
where the function $f$ can be chosen arbitrarily
(i.e.\ $\varphi$ is not unique).

Using (\ref{46}), (\ref{48}) and (\ref{50}) one now easily verifies
\begin{eqnarray}
\varphi(Y)&=&\varphi(X)-\tau L\ ;\label{liou0}\\
s_\tau\varphi(Y)&=&
\xi^\alpha\partial _\alpha \varphi(Y)-\tau \, c\ .
\label{liou1}\end{eqnarray}
(\ref{liou1}) implies that $\varphi(Y)$ plays in the deformed theory
the role of a `Liouville field' as one has
\begin{eqnarray} & &s_\tau W^0_\varphi=\tau^2 \int c\, \sqrt g R\ ;
\label{liou2}\\
& &W^0_\varphi=\int \sqrt g \left[\mbox{\small{$\frac 12$}}
g^{\alpha\beta}\partial_\alpha\varphi(Y)\cdot\partial_\beta
\varphi(Y)-\tau\varphi(Y)R\right].\quad
\label{liou3}\end{eqnarray}
This shows that for $\tau\neq
0$ the matter field independent Weyl anomaly
$\int c\sqrt g R$ may be cancelled, say at the
one-loop level, by a counterterm proportional to $\hbar W^0_\varphi$.
That is impossible in the undeformed theory (i.e. for $\tau=0$) because
$\int c\sqrt g R$ is nontrivial in $H^1({\cal S})$
and thus cannot be cancelled through
the ${\cal S}$-variation of any local functional \cite{1}
(this is reflected by
the occurrence of $\tau$ on the r.h.s. of (\ref{liou2})).

A simple example to which the above construction applies
is of course a flat target space with
$G_{\mu\nu}=
\mbox{diag}(-1,1,\ldots,1)$.
In this case the covariantly constant vectors are just constants
and thus $\Lambda(X)$ is linear in the $X^\mu$.
E.g. one can choose
$\Lambda(X)=X^1+X^2$ and $ \varphi(X)=-X^1$ % XXX
which gives % $a=-1$,
$Y^1=X^1+\tau L$, $Y^2=X^2-\tau L$ and
$Y^\mu=X^\mu$ for $\mu>2$.
Examples for curved target spaces
with covariantly constant null vectors (and nonvanishing $B_{\mu\nu}$)
can be found in \cite{ht}.

We conclude this letter with the following remarks:

1. We stress again that background charges and consistent
continuous deformations changing nontrivially
the gauge transformations exist only for
target spaces possessing covariantly constant
(null) vectors solving (\ref{16}). Namely otherwise
there are no representatives of $H^0({\cal S})$ depending
nontrivially on antifields \cite{1}.

2. Eq. (\ref{24}) does not give the most general form
of the  background charges because one can still add
a functional (\ref{1a}) to $M^0_v$.
This results in additional contributions
on the r.h.s. of Eqs. (\ref{25}) and (\ref{34}) because in general the
antibracket $(M^0_v,W^0_{(0)})$ contains matter field
dependent candidate anomalies of the form given in
Eq. (10.8) in \cite{1}. Hence, even such
anomalies can possibly be cancelled through
background charges, but, again, only in presence of
solutions of (\ref{16}).

3. Similarly, functionals
(\ref{1a}) may contribute to continuous deformations
of (\ref{1}) if there are continuous families
$G_{\mu\nu}(X,\tau)=G_{\mu\nu}(X)+\tau N_{\mu\nu}(X)+\ldots$ and
$B_{\mu\nu}(X,\tau)=B_{\mu\nu}(X)+\tau F_{\mu\nu}(X)+\ldots$
with at least one corresponding continuous set of
covariantly constant null vectors.

4. The $2d$ gravity models obtained from deformed
classical actions (\ref{36}) are new. This is evident for
left-right asymmetric actions (\ref{36}) but it holds also
for the left-right symmetric ones. Concerning the latter we
note that for instance (\ref{regact}) must not
be confused with the celebrated `dilaton action'
even though it has the same form as the latter. The crucial
difference is that (\ref{regact}) is gauge invariant at
classical level, contrary
to the dilaton action (recall that $\Lambda$ has to satisfy Eq.
(\ref{42}) and is thus not the `dilaton' as the latter is determined
by the cancellation condition for matter field dependent Weyl
anomalies, cf. \cite{strinbf} and section 11 of \cite{1}).

5. In a deformed theory $\tau$ plays the
role of a coupling constant. For $\tau\neq 0$ one might obtain
non-critical string theories because the matter field independent
Weyl anomaly might be cancelled
without introducing a new (`Liouville') field by hand (cf. above
discussion of (\ref{liou2})). This points out once more
that the deformed theory is different from the original one
and has physically interesting properties.

6. An interesting interpretation of a deformed action (\ref{36})
would arise if it would represent a quantum deformation of
a corresponding classical action (\ref{1}), i.e.
if $\tau\sim \hbar$. In that case the $\tau$-dependent
term in Eq. (\ref{37}) would be a quantum correction of order
$\hbar$ to the classical BRST-transformation. As the deformed gauge
transformations have the same algebra as the original ones,
the theory would satisfy the structural constraints of
\cite{gw}.
\medskip
\section*{Acknowledgments.}
We thank O. Lechtenfeld for pointing out to us ref. \cite{ht}.
This work was carried
out in the framework of the European Community Research Programme
``Gauge theories, applied supersymmetry and quantum gravity", with a
financial contribution under contract SC1-CT92-0789.

\end{document}